\documentclass[aps,prl,twocolumn,superscriptaddress,showpacs]{revtex4-1}

\usepackage{color}
\usepackage{graphicx}
\usepackage{dcolumn}
\usepackage{bm}
\usepackage{epsfig}                    
\usepackage{float}

\begin{document}

\title{Direct proof of unique magnetic and superconducting phases in superoxygenated high-$T_c$ cuprates}

\author{L.~Udby}
\affiliation{Nanoscience Center, Niels Bohr Institute, University of Copenhagen, DK-2100 Copenhagen, Denmark}

\author{J.~Larsen}
\affiliation{Department of Physics, Technical University of Denmark, DK-2800 Kgs. Lyngby, Denmark}

\author{N.~B.~Christensen}
\affiliation{Department of Physics, Technical University of Denmark, DK-2800 Kgs. Lyngby, Denmark}

\author{M. Boehm}
\affiliation{Institut Laue-Langevin, 38042 Grenoble Cedex 9, France}

\author{Ch.~Niedermayer}
\affiliation{Laboratory for Neutron Scattering, Paul Scherrer Institut, CH-5232 VIlligen PSI, Switzerland}

\author{H.~E.~Mohottala}
\affiliation{University of Hartford, West Hartford, Connecticut 06117, USA}

\author{T.~B.~S.~Jensen}
\affiliation{Department of Physics, Technical University of Denmark, DK-2800 Kgs. Lyngby, Denmark}

\author{R.~Toft-Petersen}
\affiliation{Department of Physics, Technical University of Denmark, DK-2800 Kgs. Lyngby, Denmark}
\affiliation{Helmholtz Zentrum Berlin f\"ur Materialien und Energie, D-14109 Berlin, Germany}

\author{F.~C.~Chou}
\affiliation{Center for Condensed Matter Sciences, National Taiwan University, Taipei 10617, Taiwan}

\author{N.~H.~Andersen}
\affiliation{Department of Physics, Technical University of Denmark, DK-2800 Kgs. Lyngby, Denmark}

\author{K.~Lefmann}
\affiliation{Nanoscience Center, Niels Bohr Institute, University of Copenhagen, DK-2100 Copenhagen, Denmark}

\author{B.~O.~Wells}
\affiliation{Department of Physics, University of Connecticut, Storrs, Connecticut 06269-3046, USA}


\begin{abstract}
We present a combined magnetic neutron scattering and muon spin rotation study of the nature of the magnetic and superconducting phases in electronically phase separated La$_{2-x}$Sr$_x$CuO$_{4+y}$, $x$ = 0.04, 0.065, 0.09. For all samples, we find long-range modulated magnetic order below $T_N \simeq Tc$ = 39~K. In sharp contrast with oxygen-stoichiometric La$_{2-x}$Sr$_x$CuO$_4$, we find that the magnetic propagation vector as well as the ordered magnetic moment is independent of Sr content and consistent with that of the 'striped' cuprates. 
Our study provides direct proof that superoxygenation in La$_{2-x}$Sr$_x$CuO$_{4+y}$ allows the spin stripe ordered phase to emerge and phase separate from superconducting regions with the hallmarks of optimally doped oxygen-stoichiometric La$_{2-x}$Sr$_x$CuO$_4$.

\end{abstract}

\pacs{74.72.Gh, 74.25.Ha, 75.25.-j,74.10.+v }
\maketitle

The many active degrees of freedom in transition metal oxides lead to intrinsic complexity with different electronic states being nearly degenerate. As a consequence nanoscale phase separation can be observed in such different materials as the CMR manganites and high-temperature superconducting (HTSC) cuprates \cite{dagotto_S309,alvarez_PRB71}. A central challenging theme is how dopant disorder influences the details of the phase separation in otherwise electronically similar systems and e.g. pins fluctuating order \cite{kivelson_RMP75}. We adress this issue by investigating the electronic properties of a HTSC system with two essentially different mechanisms of charge-carrier doping i.e. mobile oxygen ions and immobile Sr ions.\\
\indent Starting from the Mott insulating and antiferromagnetic parent compound La$_2$CuO$_4$ (LCO), replacement of La by Sr leads to superconductivity above $x=0.055$ in La$_{2-x}$Sr$_x$CuO$_4$ (LSCO) with the highest superconducting transition temperature, $T_c=38$ K at $x\simeq$ 0.15 (optimal doping) \cite{takagi_PRB40}. On the other hand, intercalation of a sufficient amount of excess oxygen 
in Sr-free samples to produce La$_2$CuO$_{4+y}$ (LCO+O) leads to even higher $T_c \simeq 42$ K \cite{wattiaux_CRASP310} and less flux-pinning \cite{chou_PC216}.
The origin of the differences in superconducting properties lies in the nature of the doping-processes: When oxygen-stoichiometric LSCO is formed by cooling through the liquid-solid phase transition at temperatures far above room temperature, a homogeneous but quenched disordered distribution of Sr on La sites is produced. By contrast, intercalated oxygen remains mobile down to much lower temperatures 
\cite{xiong_PRL76} where it tends to organise in well-ordered superstructures that can be observed in diffraction experiments \cite{wells_ZP100}, and over which there is a partial degree of control \cite{lee_PRB69,fratini_N466}. 
Combining magnetisation and muon spin rotation, we have recently discovered that even in samples containing quenched disordered Sr, intercalated oxygen facilitates optimal superconducting properties ($T^{onset}_c \simeq 40$ K and weak pinning) \cite{mohottala_NM5,mohottala_PRB78}. It does so by promoting phase separation between regions of the sample that are non-magnetic (and superconducting) and regions with magnetic order. The local magnetic fields around the muon stopping site are similar \cite{mohottala_NM5} to those of the so-called stripe ordered materials (La,Nd)$_{15/8}$Sr$_{1/8}$CuO$_4$ (LNSCO) and La$_{15/8}$Ba$_{1/8}$CuO$_4$ (LBCO) \cite{nachumi_PRB58}. 
From elastic neutron scattering (ENS) experiments on these materials it is known that the magnetic order is characterised by two incommensurate magnetic propagation vectors, corresponding to two domains of modulated antiferromagnetic order \cite{fujita_PRB70,tranquada_N375,christensen_PRL98}. An ENS study on LSCO+O, $x=0.09$ reveal similar peaks \cite{udby_PRB80}, but a systematic exposition of the nature and possible evolution of magnetic and superconducting states in LSCO+O has been lacking.\\
\indent In this Letter we present a ENS study of the magnetic properties of LSCO+O single crystals covering a broad range of Sr content, and investigate the superconducting properties using high transverse field muon spin rotation (HTF-$\mu$SR). Using neutrons as a bulk-sensitive probe of magnetism, we provide direct evidence for the identity of the magnetic phases of our LSCO+O samples in terms of propagation vector and ordered magnetic moment. Moreover, we show that these characteristics are the same as those of stripe-ordered LNSCO and LBCO. Further, we find that the superconducting penetration depth of all samples are identical within our experimental errors and of a magnitude similar to that of optimally doped oxygen-stoichiometric LSCO.\\
\indent All samples studied are the same single crystals also used in \cite{mohottala_NM5} with $x=0.04,0.065,0.09$. They were float-zone grown in an optical furnace and post-oxidised (superoxygenated) through wet-chemical methods \cite{wattiaux_CRASP310,wells_ZP100}. 
The intercalation process was stopped after a long period of oxidation, always after the sample showed a single transition of $T^{onset}_c\sim$40 K as recorded by SQUID measurements.
The ENS studies were performed at the cold triple-axis spectrometers RITA-II and IN14 at the Paul Scherrer Institute (PSI), Switzerland and the Institut Laue-Langevin, France, respectively. Both spectrometers employed elastic scattering mode with $E_i=E_f=5$ meV and 40' horizontal collimation before and after the sample. Be-filters removed higher-order contamination scattering from the monochromators. All Miller indices in this work refer to the orthorhombic $Bmab$ notation in reciprocal lattice units [rlu] based on the low temperature lattice parameters \cite{suppl}. 
The muon data were recorded at the General Purpose Surface-Muon Instrument at PSI using a high (0.3~T) transverse field after fast ($>$1~K/min) cooling, since the SC properties are known not to change with cooling rate for the investigated crystals.

\begin{figure}[!h]
\includegraphics[width=\columnwidth]{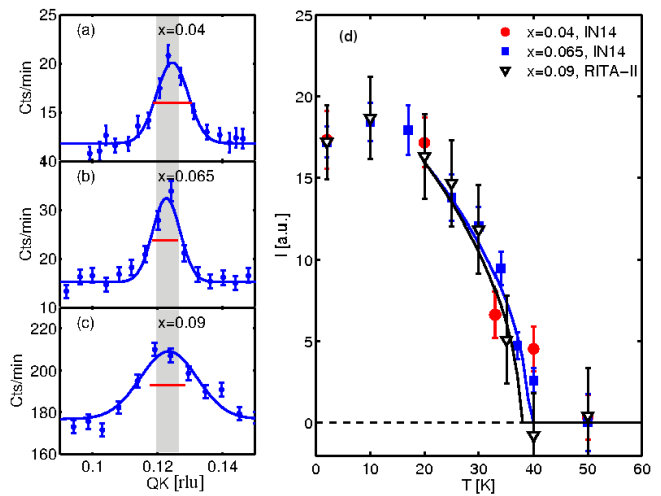}
\caption{\label{fig:sLSCO_ICpeak_Tdep}(Color online) (a-c) ENS scans in reciprocal space through the m-AFM peak at $\mathbf{Q}$=(1+$\delta_H$, $\delta_K$, 0). All data were taken at $T$ = 2~K at IN14. Solid lines are Gaussian fits and the horizontal red lines show the resolution as found by corresponding scans through Bragg peaks. The grey shaded region indicates the average peak position as described in the text. (d) Temperature dependence of the intensity of the m-AFM peaks. The intensities are scaled at 2~K. In order to locate $T_N$, power-law fits with a fixed exponent $\eta=$ 0.5 were conducted (lines).}
\end{figure}

To set the stage for LSCO+O, we start by summarising the magnetic properties of oxygen-stoichiometric LSCO: In the magnetic phase the modulation period and direction depends strongly on Sr content $x$ and a quartet of peaks are detected by ENS 
with $\delta\propto x$ \cite{yamada_PRB57} away from the anti-ferromagnetic position corresponding to modulated anti-ferromagnetic (m-AFM) order in the CuO planes. For $0.024 \lesssim x \lesssim 0.055$ the spin structure is rotated i.e modulated diagonally with respect to the Cu-O bonds \cite{fujita_PRB65, wakimoto_PRB61,matsuda_PRB62}. For $x > 0.055$ the modulation is parallel to the Cu-O bonds with incommensurability saturating at $\delta \simeq 1/8$ for $x \simeq 1/8$ \cite{yamada_PRB57}. Long-range magnetic order with correlation length $\xi >$ 100 \AA{} is only found for $x \simeq$ 1/8.
In striking contrast with these characteristics of LSCO, Figure  \ref{fig:sLSCO_ICpeak_Tdep} shows several key results of our study:
In all the investigated superoxygenated LSCO+O samples through the Sr doping range $x=0.04-0.09$ at $T\sim$2~K, we have observed a quartet of peaks by ENS at the \emph{same} positions. The peaks at $\mathbf{Q}=(1+\delta_H,\delta_K,0)$ are compared in the left panel of Figure \ref{fig:sLSCO_ICpeak_Tdep}.
We find for all $x$ that the peaks are located $\delta_H\sim\delta_K\sim\delta=0.123\pm0.004$ away from the antiferromagnetic point. This corresponds to m-AFM with periodicity of $8.1(3)$ unit cells parallel the Cu-O bonds as is also found in oxygen-stoichiometric LSCO $x\simeq 1/8$. 
That the incommensurability $\delta$ and the modulation direction is always the same in LSCO+O regardless of Sr doping $x$ is however opposed to what is observed in oxygen-stoichiometric LSCO.
The peaks of LSCO+O  are sharp and instrumentally resolved for $x<0.09$ as seen in the left panel of Figure \ref{fig:sLSCO_ICpeak_Tdep}. For  $x=0.09$ there is however a $\sim$30\% broadening which we previously found to result from the finite size of the m-AFM domains in the sample \cite{udby_PRB80,udby_NIMA634}. 
These domains are, however, at least $300-400$~\AA{} for all $x$.
This periodicity and long correlation lengths of the m-AFM signal are similar to those of the zero-field magnetic signal observed in LCO+O \cite{lee_PRB60,khaykovich_PRB66}, oxygen stoichiometric LSCO with $x\simeq 1/8$ and the parallel stripes found in LNSCO and LBCO \cite{tranquada_N375,christensen_PRL98,fujita_PRB70}. The spin correlation lengths in our LSCO+O samples are however much larger than in oxygen stoichiometric LSCO samples with comparable Sr content \cite{yamada_PRB57}. 
The temperature dependence of the m-AFM peak intensity for LSCO+O is shown in the right panel of Figure \ref{fig:sLSCO_ICpeak_Tdep}. It follows the same power-law dependence for all  $x=0.04-0.09$ with transition temperature $T_N=39(3)$~K \footnote{We have too few points in the ENS data to fit a proper $T_N$ for the small $x=0.04$ crystal, but $T_N\sim40$K is consistent with the $\mu$SR data as discussed later in the text} which is also in contrast to oxygen stoichiometric LSCO where the intensity does not follow the usual power-law dependence for the lowest dopings \cite{wakimoto_PRB60} and $T_N$ is much smaller and varies in the same Sr doping range \cite{hirota_PC357}.

\begin{figure}[h] 
       \includegraphics[width=\columnwidth]{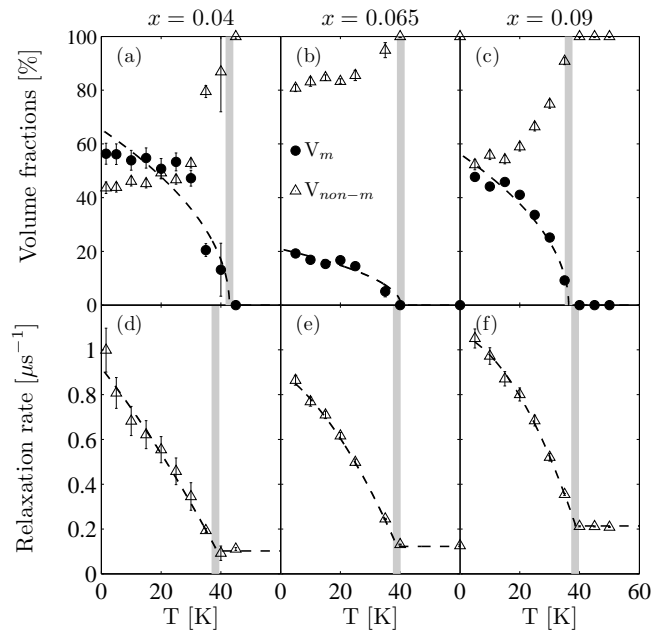}
       \caption{\label{fig:HTF}\small Results of HTF-$\mu$SR experiments. If not visible, the statistical errors are smaller than the datapoint markers. (a-c) Temperature dependence of the magnetic and non-magnetic volume fractions of the sample. Dashed lines are guides to the eye and grey shaded regions mark $T_N$.
(d-f) Temperature dependence of the spin relaxation rate for muons stopping in the superconducting regions of the sample. Following  \cite{sonier_PRL72} the dashed lines are fits to the function $\sigma_0(1-aT-bT^2)$ for $T<T_c$. Grey shaded regions mark $T_c$.}
\end{figure}

In order to find the magnitude of the ordered magnetic moments from the ENS data we need knowledge of the magnetic volume fractions which can be provided by $\mu$SR experiments. 
The muons stop at specific lattice positions and provide a random sampling of the internal field distribution both in the magnetic volume fraction and the vortex state of the superconducting volume fraction.
All $\mu$SR data presented in this Letter were fitted with a three-component model for the assymetry following the procedure outlined in \cite{ansaldo_PRB40}. Two of the components are temperature dependent and related to the sample. A third component models the background originating from e.g. muons stopping in the cryostat walls or sample holder and is assumed to be temperature independent. For details see the supplementary information \cite{suppl}.
The first temperature dependent component models the muons which are rapidly relaxing, i.e. they are being depolarised by the ordered moments in the magnetic volume fraction of each sample for which the derived temperature dependence is shown in the top panels of Figure \ref{fig:HTF}. We note that the magnetic volume fractions begin to grow at the same temperatures at which ENS reveals the onset of m-AFM order, see Figure \ref{fig:sLSCO_ICpeak_Tdep}(d), indicating that truly static magnetic order sets in below $T_N$. The second temperature dependent component is slowly relaxing and originates from the non-magnetic part of the sample. Its temperature dependence is also shown for each sample in the top panels of Figure \ref{fig:HTF}. A slight decrease in precession frequency of this component \cite{suppl} marks the superconducting onset transition at $T_c$. At base temperature we assume that all of this component originates from the flux-line lattice in the superconducting volume of the sample.
The temperature dependence of the relaxation rate in the non-magnetic volume is shown in Figure \ref{fig:HTF}. For all samples we observe a similar temperature dependence with $\sigma(T \to 0)\sim 0.9\ \mu{\rm s}^{-1}$. This value is the expected relaxation value for a superconducting volume with a penetration depth of at least $\lambda\sim 1500$ \AA{} \cite{Uemura_PRB38} in an optimally doped LSCO sample with a rigid 3D vortex lattice. It is seen from these data that the relaxation rate increases below 40~K in all samples, coinciding with the superconducting transition temperature $T_c=39(1)$~K. We have confirmed this by AC susceptibility measurements, regardless of the cooling rate. The magnetic and superconducting volume fractions and their transition temperatures are compiled in Table \ref{tab:magSC}.  We note that the magnetic and superconducting transition temperatures coincide for all $x$. 

\begin{table}[!h]
\begin{ruledtabular}
\begin{tabular}{|l|l|l|l|l|}
x       & $V_m $[\%] &  $V_{SC}$[\%] & $T_N$[K] & $T_{c}$[K]\\
\hline
0.04    & 56(4)       & 44(1)   & $\sim$40    &   39(1)  \\
0.065   & 19(1)       & 81(1)   & 39(1)    &   38.7(7) \\
0.09    & 47(1)       & 53(1)   & 38(2)    &   37.1(5) \\
\end{tabular}
\end{ruledtabular}
\caption{
Collected $\mu$SR results for the magnetic and superconducting base temperature volume fractions, $V_m$ and $V_{SC}$, respectively, of each sample. $T_N$ and $T_c$ are determined from ENS (Fig. \ref{fig:sLSCO_ICpeak_Tdep}) and $\mu$SR (Fig. \ref{fig:HTF} (d-f)), respectively.}
\label{tab:magSC} 
\end{table}
\begin{table}[!h]
\begin{ruledtabular}
\begin{tabular}{|l|l|l|l|l|}
x       &  $m$ [g] & $I_{IC}[\frac{\rm{cts}}{\rm{min}\cdot \rm{g}}$] & $I_{200}[\frac{\rm{cts}}{\rm{min}\cdot \rm{g}}]$ & $A=\frac{I_{IC}}{I_{200}\cdot V_m}$\\
\hline
0.04    & 0.035  &  2.9(5)   & 3.7(3)$\cdot10^4$    & 1.4(3)$\cdot10^{-4}$\\
0.065   & 0.091  &  2.0(3)   & 6.4(2)$\cdot10^4$    & 1.6(3)$\cdot10^{-4}$\\
0.09    & 0.415  &  1.7(3)   & 2.0(1)$\cdot10^4$    & 1.8(3)$\cdot10^{-4}$\\
\end{tabular}
\end{ruledtabular}
\caption{ 
Mass ($m$), mass-normalized IC AFM and nuclear Bragg peak intensities, I$_{IC}$ and I$_{200}$, respectively. The
intensities derive from Gaussian fits to data taken under identical experimental conditions at IN14. The
last column shows I$_{IC}$ normalized by the product of I$_{200}$ and the magnetic volume fraction $V_m$.}
\label{tab:sLSCO}
\end{table}

\begin{figure}[!h]
\includegraphics[width=\columnwidth]{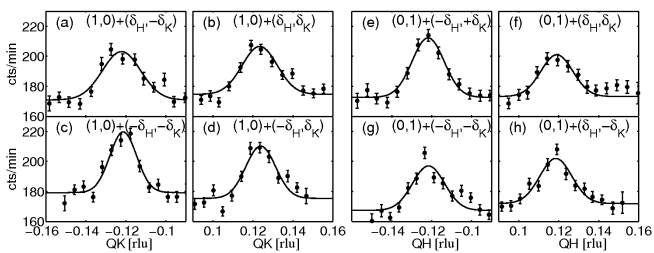}
\caption{\label{fig:sLSCO9_ICpeaks}(a-h) Rocking curve scans through the IC AFM peaks of LSCO+O $x=0.09$. All lines represent Gaussian fits.}
\end{figure}

We now return to the derivation of the magnetic moment from the ENS data based on the aquired knowledge of the magnetic volume fractions.
Table \ref{tab:sLSCO} shows the integrated, mass-normalised peak areas of the m-AFM peaks from the ENS data, and it it seen that the intensity varies substantially, probably due to differences in both magnetic volume fraction and sample mosaic details. Hence, we normalise the m-AFM peaks to the integrated area of a Bragg peak and divide by the magnetic volume fraction to obtain the constant $A$ listed in Table \ref{tab:sLSCO}. It is seen that  $A$ is the same for all LSCO+O samples within errors, thus implying they have the same ordered magnetic moment if the same model for the magnetic order can be assumed.
We thus turn to the specifics of the symmetry-related peaks in order to motivate a model for the spin-structure.
For all LSCO+O samples we have observed peaks at the same positions $\mathbf{Q}_{m}=(1\pm\delta_H,0\pm\delta_K,0)$ and $(0\pm\delta_H,1\pm\delta_K,0)$ indicating a similar spin structure.
For the $x=0.09$ sample, full scans at all above mentioned positions were performed and the data fitted to Gaussian lineshapes as shown in Figure \ref{fig:sLSCO9_ICpeaks}. The intensities and widths of all the m-AFM peaks are found to be the same within two standard deviations as is also observed for the spin stripes in LNSCO \cite{christensen_PRL98} and LBCO \cite{fujita_PC481}.
We therefore assume, analogously to LNSCO, that the m-AFM peaks in LSCO+O in one direction can be represented by a simple collinear spin stripe model and the other set of peaks in the quartet are generated by 90$^\circ$ rotation beween alternating CuO planes.
For each CuO plane we consider a $8\times2$ Cu-site unit cell $[\uparrow \downarrow \uparrow \cdot \downarrow \uparrow \downarrow \cdot\ ;\ \downarrow \uparrow \downarrow \cdot \uparrow \downarrow \uparrow \cdot]$  where the moments are lying in the CuO plane with angle $\beta$ with  $\mathbf{Q}_{m}$. Then the magnetic structure factor is given by
$\left| F_{m}\right|^2 = p^2 f_{m}^2\mu^2\sin^2\beta | \tilde{F}_{m} |^2$.
For $S$=1/2 spins we have $p=0.2696\cdot 10^{-14}$~m, and the form factor for Cu$^{2+}$ and geometrical structure factor take the values $f_{m}=0.90(5)$ \cite{shamoto_PRB48} and $| \tilde{F}_{m}|^2 =93.25$ at the m-AFM points. Details of this and the following calculations are shown in the supplematary material \cite{suppl}.
Based on the experimentally determined factor $A$ in Table \ref{tab:sLSCO} we find the ordered moment in units of Bohr magnetons to be 
\begin{equation}\label{eq:mu2}
\mu = \sqrt{\left| F_{m}\right|^2/p^2f_{m}^2\sin^2{\beta}\left|\tilde{F}_{m}\right|^2} = 0.10(2)\sqrt{C}/\sin{\beta}
\end{equation}
 Assuming that the spins are weakly correlated between neighboring CuO planes, the vertical resolution correction gives $C \simeq 1.4$ \cite{suppl} and we obtain $\mu= 0.12(2)\mu_B$ for spins oriented along [010] ($\sin \beta =0.99$) as in La$_2$CuO$_4$ \cite{vaknin_PRL58} and $\mu=0.17(3) \mu_B$ for spins oriented along [110] ($\sin \beta = 0.7(1)$), as observed in LNSCO \cite{christensen_PRL98}.  
If the scattering intensity is approximately constant along $c^\ast$ (scattering rods) due to e.g. twinning
as seen in LCO+O \cite{lee_PRB60} we have  $C \simeq 2.7$ and the quoted ordered moments must be corrected by a numerical factor $\sim 1.4$. 
In the absence of experimental information about the $c$-axis magnetic correlations, we restrain ourselves to the conclusion that the ordered magnetic moments in the magnetic volume fractions of LSCO+O are of the same order of magnitude as the those determined for LCO+O ($\mu=0.15(5) \mu_B$ \cite{lee_PRB60}) and stripe ordered LNSCO ($\mu=0.10(3) \mu_B$ \cite{tranquada_PRB54}). 

We summarise our results of LSCO+O through a range of Sr dopings to conclude that superoxygenation facilitates the same type of long-range m-AFM order, with same periodicity $\delta\sim1/8$, moment and transition temperature $T_N=39(3)$ K within errors. This is in contrast to oxygen stoichiometric LSCO \cite{hirota_PC357,wakimoto_PRB61,yamada_PRB57}. 
Furthermore, the magnetic transition temperatures of the studied LSCO+O crystals are the same within errors whether determined using the local $\mu$SR probe or bulk-sensitive neutron scattering. 
This is also in contrast to oxygen stoichiometric LSCO where the transition temperature as observed by $\mu$SR is significantly lower than the one observed by neutron scattering due to gradual freezing of the moments \cite{julien_PB329}. 
These observations are evidence of the existence of a single and long-range ordered m-AFM phase thoughout the Sr doping range $0.04\leq x\leq 0.09$ in LSCO+O which is similar to the striped systems LBCO \cite{fujita_PC481,fujita_PRB70} and LNSCO \cite{christensen_PRL98,tranquada_N375,chang_PRB78,wen_PRB78}. 
 Since our LSCO+O crystals stay orthorhombic at low temperatures, the stripe-like magnetic order is not pinned by the strong ordering field of the low temperature tetragonal (LTT) phase as in LBCO and LNSCO\cite{fujita_PRB70,tranquada_N375}. However, the commensurate nature of the ordering still implies a strong coupling to the lattice. We expect that the difference here is that since LSCO+O  has a weaker, random disordering field from the Sr dopants it also does not require the stronger lattice ordering field associated with the LTT phase. This picture is consistent with our observation that the LSCO+O sample with the highest Sr content $x=0.09$ has slightly reduced magnetic correlation length.

Regarding the superconducting phase of LSCO+O, the transition temperature $T_c=39(1)$~K is the same within errors and coinciding with the magnetic transition temperature for all samples in contrast to $T_c \propto x$ in oxygen-stoichiometric LSCO \cite{yamada_PRB57}. $T_c$ is also not suppressed as in the anomalous 1/8 state of LNSCO \cite{crawford_PRB44} and LBCO \cite{moodenbaugh_PRB38} suggesting  phase separation rather than competition between the two phases in the same areas of the sample.
Furthermore the penetration depth is similar in the superoxygenated system thoughout the investigated Sr range and has value corresponding to that of optimally doped LSCO. This is in contrast to oxygen stoichiometric LSCO where 
the penetration depths for superconducting samples increase with $x$ \cite{uemura_PRL62}. The similarity of the temperature dependence of the relaxation rate in LSCO+O indicates that the superconducting gap symmetry is similar thoughout the Sr doping range.\\
\indent These observations prove that a long-range electronic phase separation occurs in LSCO+O between a 1/8 stripe-like magnetic phase and a superconducting phase which is similar to optimally doped LSCO.
 A recent report on  pressure studies of LBCO \cite{guguchia} has $\mu$SR and magnetisation data similar to our earlier work \cite{mohottala_NM5} revealing 
phase separation between stripe-like magnetism and superconductivity. Despite these striking similarilies between LSCO+O and pressurised LBCO,
the latter seems to favor an underdoped, reduced $T_c$ superconducting phase perhaps by depinning the charge order \cite{huecker_PRL104}. 
This differs from the superoxygenated samples we describe here where the separate phases appear to have different effective charge densities.\\
\indent We thank Brian M. Andersen for helpful discussions. This work was supported by the Danish Agency for Science Technology and Innovation under the Framework Programme on Superconductivity and the Danish Research Council FNU through the instrument center DANSCATT. Work at the University of Connecticut was supported by the U.S. Department of Energy under Contract No. DE-FG02-00ER45801. This work is based on experiments performed partly at the Institut Laue-Langevin, Grenoble, France and partly atthe  Swiss spallation neutron source SINQ, Paul Scherrer Institute, Villigen, Switzerland. 
\bibliographystyle{apsrev}
\bibliography{references_PRL}

\end{document}


\section*{Supplementary Material for: Direct proof of unique magnetic and superconducting phases in superoxygenated high-$T_c$ cuprates}

\subsection*{Crystal lattice details}
The lattice constants as derived from neutron scattering measurements are provided in Table \ref{tab:latticepar}.
\begin{table}[!h]
\begin{ruledtabular}
\begin{tabular*}{6cm}{|l|l|l|l|}
x       & $a$[\AA] &  $b$ [\AA] & $c$[\AA] \\
\hline
0.04    & 5.32(3)       & 5.38(3)   & 13.19(1)    \\
0.065   & 5.32(2)       & 5.36(3)   & 13.15(3)    \\
0.09    & 5.33(3)       & 5.34(3)   & 13.14(2)    \\
\end{tabular*}
\end{ruledtabular}
\caption{Low temperature lattice parameters for our three samples, values found in \cite{udby_thesis}.}
\label{tab:latticepar} 
\end{table}

Since the orthorhombic distortion is small for all crystals ($2(a-b)/(a+b)< 1$\% ) and since $\delta_H\approx\delta_K$, we compare the incommensurability of the magnetic peaks in pseudo-tetragonal units where $b_T=a_T=(a+b)/(2\sqrt{2})$.

\subsection*{Intensity corrections of neutron scattering data}
In order to optimise the signal/background ratio in the neutron scattering experiments and make sure we record only scatteing from static (or close to static) magnetic order in the data, we have chosen to perform the experiments on cold triple-axis spectrometers which have an energy-resolution better than  $\Delta E=0.2$ meV.\\
The measured integrated intensity of the elastic neutron scattering peaks which are printed in Table II of the Letter are obtained from Gaussian fits to scans which are close to rocking curves. 
Theoretically the integrated intensity of a peak recorded by a rocking curve at a given neutron wavelength $\lambda$ can be written \cite{shirane}
\begin{equation}\label{eq:I}
I_{hkl}^{\rm cal} = \Phi_\lambda \frac{(2\pi)^3 N}{\mathcal{V}_0 (\sin{2\theta_{hkl})_\lambda}}\lambda^3\left|F_{hkl}\right|^2
\end{equation}
where $\Phi_\lambda$ is the neutron flux at the given wavelength, $2\theta_{hkl}$ is the scattering angle and $\left|F_{hkl}\right|^2$ the structure factor for the given reflection. $N$ is the number of unit cells of volume $\mathcal{V}_0$ in the sample. 

Let us assume, in analogy with LNSCO \cite{christensen_PRL98}, that the modulated AFM (m-AFM) peaks in LSCO+O in one direction e.g. $\mathbf{Q}_{IC}=(1-\delta_H,\delta_K,0)$ can be represented by a simple collinear spin stripe model where the moments $\mu$ on the Cu-atoms are lying in the CuO plane with angle $\beta$ with  $\mathbf{Q}_{IC}$. 
The magnetic structure factor is given by
\begin{equation}\label{eq:Fmag}
\left| F_{hkl}\right|^2 = p^2 f_{hkl}^2\mu^2\sin^2\beta \left| \tilde{F}_{hkl}\right|^2
\end{equation}
where $pf(\mathbf{Q}_{hkl})$ is the neutron magnetic scattering length with $p=0.2696\cdot 10^{-14}$~m for $S$=1/2 spins and $f(\mathbf{Q}_{hkl})$ the $\mathbf{Q}$-dependent form factor for Cu$^{2+}$ \cite{shamoto_PRB48} which takes the value $f(\mathbf{Q}_{m}) \equiv f_{m}=0.90(5)$ at $Q_m=1.2\pm 0.2$\AA{}$^{-1}$ of the m-AFM points. We consider a $8\times2$ Cu-site unit cell $[\uparrow \downarrow \uparrow \cdot \downarrow \uparrow \downarrow \cdot\ ;\ \downarrow \uparrow \downarrow \cdot \uparrow \downarrow \uparrow \cdot]$ for which the geometrical structure factor is $| \tilde{F}_{hkl}| = \left(1+e^{i2\pi h}+e^{i4\pi h}\right)\left(1-e^{i8\pi h}\right)\left(1-e^{i\pi(k-h))}\right)$ using orthorhombic Miller indices. At the m-AFM points we get $| \tilde{F}_{IC}|^2 =93.25$. The other set of peaks in the quartet comes from 90$^\circ$ rotation beween alternating CuO planes in this model.

The ratio of the integrated intensity of an m-AFM peak in the magnetic volume $\mathcal{V}_m$ and a nuclear Bragg peak integrated intensity is comparable to the measured intensity ratio with appropriate corrections for vertical spectrometer resolution $\Delta q_z$ and extinction $B$. Therefore we write
\begin{equation}\label{eq:A}
\tilde{A} \equiv I^{cal}_{m} / I^{cal}_{200}= (I^{mea}_{m}/\mathcal{V}_m) \cdot (nc^\ast/\Delta q_z)/(I^{mea}_{200}\ B) = A C
\end{equation}
where $A=I^{mea}_{m}/(\mathcal{V}_m \cdot I^{mea}_{200}) =1.6(3)\cdot 10^{-4}$ is experimentally determined and printed in Table II of the Letter. The vertical resolution $\Delta q_z\sim 0.18$~\AA{}$^{-1}$ is determined from the instrument geometry and $n$ is a correction factor for the distribution of the scattering intensity out of the scattering plane:
If we assume that the spins are weakly correlated between neighboring CuO planes and the scattering intensity is correspondingly sinosoidally modulated along $c^\ast$, as seen in LNSCO \cite{tranquada_PRB54}, we have $nc^\ast=0.25$~\AA{}$^{-1}$ and get $C \simeq 1.4$. If the scattering intensity is constant along $c^\ast$ in the first Brillouin zone (scattering rods), as seen in LCO+O \cite{lee_PRB60}, we have $nc^\ast=0.48$~\AA{}$^{-1}$ and get $C \simeq 2.7$.
 The extinction coefficient for the (200) second order Bragg peak can be neglected ($B\approx 1$) according to the following arguments: The strongest intensity from any of our crystals at this reflection was 1.2$\cdot 10^4$ n/s. By McStas \cite{willendrup_PB350}
Monte Carlo neutron ray tracing based on the model of IN14 by Emmanuel Farhi (Computing for Science, Institute Laue-Langevin) we have simulated the incident flux at the sample position to be 9.4$\cdot 10^{6}$ n/s*cm$^2$ for 20 meV (second order) neutrons. Since the sample cross-section area is approximately 0.1~cm$^2$ the total reflectivity from the crystal in this scattering condition is of the order of 1\%, making a 2nd order correction as extinction negligibly small.\\
\indent Combining (\ref{eq:I})-(\ref{eq:A}) we get
\begin{equation}
\left| F_{m}\right|^2 = A C \left|F_{200} \right|^2 \frac{\Phi(2\text{\AA\ })}{\Phi(\text{4\AA\ })}\frac{N\ \mathcal{V}_{m}}{N_{m}\ \mathcal{V}_0}\frac{(\sin{2\theta_{m}})_{\text{4\AA}}}{(\sin{2\theta_{200}})_{\text{2\AA}}}\left(\frac{2\text{\AA\ }}{4\text{\AA\ }}\right)^3
\end{equation}
The scattering angles 2$\theta$ for the m-AFM peaks and second order scattering off (200) are similar so the geometrical intensity correction for scattering angle (the so-called Lorentz factor, see e.g. \cite{shirane}) cancels out. The flux ratio is 0.47 by time-of-flight measurements at the sample position of IN14.
Furthermore we calculated $\left|F_{\rm 200}\right|^2 = 93$~barns based on atomic positions given in \cite{rial_PC254} and found that it does not vary more than 1\% between superoxygenated powder samples with $x=0-0.09$. Hence we exclude variations in the Bragg peak intensity as a source of uncertainty and the moment is given by
\begin{equation}\label{eq:mu2}
\mu = \sqrt{\frac{\left| F_{m}\right|^2}{p^2f^2_{m}\sin^2{\beta}\left|\tilde{F}_{m}\right|^2}} = 0.10(2)\sqrt{C}/\sin{\beta}
\end{equation}
As mentioned above realistic values for the vertical resolution correction $C$ lies in the range 1.4-2.7 yielding a correction factor of 1.2-1.7 to the value of 0.1 of the moment. If the spin direction is roughly along the [010] axis of the crystal structure as in the parent compound LCO \cite{vaknin_PRL58} we have $\sin{\beta}$=0.99 . If however the spins are roughly along the [110] direction as one model of LNSCO predicts \cite{christensen_PRL98}, then $\sin{\beta}=0.7(1)$. 

\subsection*{Details of the fitting model for the $\mu$SR data}
The $\mu$SR experiments were performed at the General Purpose Surface-Muon instrument at the Paul Scherrer Institute, Switzerland which has a time resolution of $1\cdot 10^{-8}$ s corresponding to an energy resolution of $\Delta E=4\cdot10^{-4}$ meV. The data were analysed with the free software package MUSRFIT \cite{musrfit} and fit to the following expression  
\begin{equation}\label{eq:assymetry}
\begin{align}
	G_{z}\left( t \right) &= 
A_{1} \exp{\left( - \frac{1}{2} \left( \sigma_{1} t_{} \right)^2 \right)} \cos \left( \omega_{1}t + \delta \right) \\ 
	&+ A_{2} \left[ \frac{\sigma_{-}}{\sigma_{-}+\sigma_{+}} \exp{ \left( - \frac{1}{2} \left( \sigma_{-} t_{} \right)^2 \right)} \right. \\ 
         &\quad \quad \times \left\{ \cos \left( \omega_{2} t + \delta \right) + \sin \left( \omega_{2} t + \delta \right) \textrm{Erfi}\left( \frac{\sigma_{-} t}{\sqrt{2}} \right) \right\} \\ 
	&\quad \quad+ \left. \frac{\sigma_{+}}{\sigma_{-}+\sigma_{+}} \exp{ \left( - \frac{1}{2} \left( \sigma_{+} t_{} \right)^2 \right)} \right. \\ 
        &\quad \quad \left.\times  \left\{ \cos \left( \omega_{2} t + \delta \right) - \sin \left( \omega_{2} t + \delta \right) \textrm{Erfi}\left( \frac{\sigma_{+} t}{\sqrt{2}} \right) \right\} \right] \\ 
        &+ A_{3} \exp{\left( - \frac{1}{2} \left( \sigma_{3} t_{} \right)^2 \right)} \cos \left( \omega_{3}t + \delta \right) 
\end{align}
\end{equation}
where $\textrm{Erfi}\left(x\right) = \frac{2}{\sqrt{\pi}} \int_{0}^{x} e^{t^{2}} \rm{d} t$ is the imaginary error function. The raw data taken at $T$=5~K and their fits are shown in panels (a)-(c) of Figure \ref{fig:muSR5K} along with the frequency distributions in panels (d)-(f). The expression (\ref{eq:assymetry}) has three components $\left\{A_{1},A_{2},A_{3}\right\}$ where the first component models the fast relaxing magnetic part of the sample where muons precess with the frequency corresponding to the external field. The second component is the non-magnetic part of the sample which at base temperature equals the superconducting part of the sample. It is modeled by a slowly relaxing cosine oscillation convoluted with a skewed Gaussian function. The skewness $\eta = \sigma_{-} / \sigma_{+}$ is a result of the asymmetric field distribution of the flux line lattice in the superconductor as shown in panels (d)-(f) of Figure \ref{fig:muSR5K}. The temperature below which superconductivity occurs is also observed as a slight decrease in the frequency of the second component, see panels (g)-(h) of Figure \ref{fig:muSR5K}. The third component precesses with the external field frequency and models the very slowly relaxing background from the sample environment such as e.g. the sample holder and cryostat walls. For all temperatures the boundary condition $A_1+A_2+A_3=1$ was used in the fits, yielding $A_1=0$ above $T_N\sim 40$~K. 

\begin{figure}[!h] 
       \includegraphics[height=8.25cm]{HTF_RevisedModelRawData5K.eps}
       \includegraphics[height=8.6cm]{SkewedGssFourier_oneYlabel.eps}
       \includegraphics[height=8.5cm]{Frequency.eps}
       \caption{\label{fig:muSR5K}\small (Color online) (a-c) High-transverse field muSR data recorded at 5 K. Raw data (black points) are plotted in the rotating reference frame (frequency 38.5 MHz). Red lines are fits of the asymmetry to the three component model as described in the text. (d-f) The frequency distributions (Fourier power spectra) corresponding to the fits in panels (a-c). The grey line marks the precession frequency of the external field. (g-i) Temperature dependence of the frequency of the slowly relaxing component.}
}
\end{figure}

The low-temperature assymetry fractions with respect to the total assymetry signal (including background) are shown in Table \ref{tab:assymfracs}. 
As expected the fraction of the total signal coming from the the sample environment (holder etc.) is largest for the smallest sample ($x=0.04$).
\begin{table}[!h]
\begin{ruledtabular}
\begin{tabular}{|l|l|l|l|l|}
x       & $m$ [g] & $A_1/A_{tot} $[\%] &  $A_2/A_{tot}$[\%] & $A_3/A_{tot}$[\%]\\
\hline
0.04    & 0.035 &    44(3)       & 35(1)   & 21 \\
0.065   & 0.091 &    18(1)       & 76(1)   & 6  \\
0.09    & 0.415 &    45(1)       & 50(1)   & 5  \\
\end{tabular}
\end{ruledtabular}
\caption{Sample Sr content and mass are shown in the first two columns and fits to the three components of the muon assymmetry at 5 K sre shown in the last three columns.}
\label{tab:assymfracs} 
\end{table}

\bibliographystyle{apsrev}
\bibliography{references_PRL}